\documentclass[%
 reprint,
 amsmath,amssymb,
 aps,nofootinbib,longbibliography   
]{revtex4-1}
\usepackage{bm}
\PassOptionsToPackage{linktocpage}{hyperref}
\usepackage[hyperindex,breaklinks]{hyperref}
\usepackage{enumitem}
\usepackage{slashed}

\renewcommand{\theta}{\vartheta}

\usepackage{comment}

\usepackage{amsmath}
\usepackage{array}
\usepackage{mathtools}
\usepackage{xcolor}

\usepackage{etoolbox}
\smallskipamount
\begin{document}

\title{Minimal SU(5) theory on the edge: the importance of being effective}

\author{Goran Senjanovi\'c}
\email{goran.senjanovic@physik.uni-muenchen.de}
\affiliation{
Arnold Sommerfeld Center, Ludwig-Maximilians University, Munich, Germany
}
\affiliation{
International Centre for Theoretical Physics, Trieste, Italy
}
\author{Michael Zantedeschi}
\email{zantedeschim@sjtu.edu.cn}
\affiliation{Tsung-Dao Lee Institute and School of Physics and Astronomy,
Shanghai Jiao Tong University, Shanghai, China}

\begin{abstract}
It is well known that the minimal renormalizable $SU(5)$ grand unified theory is ruled out: it predicts same masses of down quarks and charged leptons, the gauge couplings do not unify and neutrinos are massless. We show here that all this can be cured simultaneously by the addition of higher-dimensional effective operators. However, the theory lives on the edge since the unification scale turns out as low as roughly $10^{14}\,\rm GeV$, 
threatening proton longevity. If the lower bound on the proton lifetime was to be increased by an order of magnitude, the 
usual desert in energies between the weak and unification scales would be populated. We also revisit two minimal extensions of this theory that offer a dynamical seesaw origin of neutrino mass, and discuss the resulting consequences.
\end{abstract} 

\maketitle

\section{Introduction} 
By unifying electro-weak and strong interactions in a single compact gauge group, grand unified theories (GUTs) explain the mystery of charge conjugation in nature and in turn predict the existence of magnetic monopoles~\cite{Polyakov:1974ek,tHooft:1974kcl}. Moreover, they imply proton decay and provide a rough estimate of its lifetime, tantalizingly close to the present experimental limits~\cite{Super-Kamiokande:2020wjk}. 

While there are many models on the market, the original $SU(5)$ theory of Georgi and Glashow~\cite{Georgi:1974sy} stands out due to its simplicity and predictivity. Actually, in its simplest renormalizable form, it is predictive enough to be ruled out by experiment due to two fundamental failures: it predicts equal masses for  charged leptons and down quarks, and gauge couplings do not unify - $\alpha_1$ meets $\alpha_2$ too early, contrary to what  seemed originally~\cite{Georgi:1974yf}. Furthermore, neutrinos end up being massless, but that can be cured by adding fermion singlets - without altering the gauge structure of the theory - which through the seesaw  mechanism~\cite{Minkowski:1977sc,Mohapatra:1979ia,Yanagida:1979as,GellMann:1980vs,Glashow:1979nm}, naturally provide small neutrino masses. 

There are two distinctive ways of potentially salvaging the theory. One is to add more fields, and take a road of model building, a field interesting in itself - something, however, we will not pursue here.
Alternatively, one can employ higher-dimensional effective operators to correct the bad fermion mass relations~\cite{Ellis:1979fg} and provide non-vanishing neutrino masses~\cite{Weinberg:1979sa},  
and even ensure gauge coupling unification~\cite{Shafi:1983gz}. 

We follow here the latter road and reanalyze the minimal $SU(5)$ theory,
augmented by dimension $d=5$ operators. A rather low unification scale and seemingly too rapid proton decay, 
prompted a belief that the theory was ruled out, repeated often over the years.
The point, however, is that proton could be stable in the limit of the tiny third generation CKM angle going to zero~\cite{Nandi:1982ew}, in which case, proton lifetime would be enhanced by some five orders of magnitude. This has been discussed at length~\cite{Berezinsky:1983va,Bajc:2002bv,BPS,FileviezPerez:2004hn,Dorsner:2004xa}, but still, no serious attempt was made to verify the validity of the  theory. After all, it was failing on three fronts as we mentioned above, and it all indicated that it could not survive experimental challenges. And so, the theory kept being sentenced to death by experts, including the present authors (especially one of them~\cite{Senjanovic:2009kr}). By today it became a gospel - the most recent review of proton decay~\cite{Ohlsson:2023ddi} even cites an extension of this theory as the minimal one. 

We show, however, that there exists a region of parameter space where this theory is still phenomenologically viable. It lives on the edge though - an experimental improvement on proton lifetime limit by a factor 10 (20) would point towards the existence of a new light scalar state, the color octet or the weak triplet, below  $100\,\rm TeV$ ($10 \,\rm TeV$) energies. 
In other words, the infamous desert at energies between the weak and the GUT scales would not exist in this case, and 
further improvements on proton lifetime limits 
could finally rule out the remaining region of the parameter space. Until that happens, we believe that desires to bury what is arguably the minimal grand unified theory, should be put to rest. 

The rest of this work, devoted to demonstrating our claim, is organized as follows. In the next Section, the central features of the minimal $SU(5)$ theory are  summarized. In Sec.~III, its gauge-coupling running is analyzed, with particular focus on the impact of particle thresholds, as well as the effect of $d=5$ operators on unification conditions. Sec.~IV discusses the flavour of proton decay, and its constraints on the unification scale. In Sec.~V, we revisit the predictions of two minimal models that generate neutrino mass through renormalisable interactions. Our findings are summarized in Sec.~VI.

\section{The minimal $SU(5)$ theory}
The minimal $SU(5)$ theory contains three fermion generations of the following representations
\begin{equation}
\label{fermions}
    \overline 5^{i}_{\rm F},\quad 10^{i}_{\rm F},\qquad i=1,2,3\,,
\end{equation}
and the adjoint and fundamental Higgs scalar representations
\begin{equation}
\label{higgs}
    24_{\rm H},\quad 5_{\rm H}\,,.
\end{equation}
The adjoint Higgs is responsible for the GUT scale symmetry breaking $M_{\rm GUT}$, whereas the fundamental one provides the electro-weak scale $M_{\rm W}$, with 
\begin{equation}
	\label{eq:vevs}
\begin{split}
\langle 24_{\rm H}\rangle &= M_{\rm GUT} \,{\rm{diag}}\left(1,1,1,-3/2,-3/2 \right); \\
 &\langle 5_{\rm H}\rangle^T = M_{\rm W} \left(0,0,0,0,1\right).
 \end{split}
\end{equation}
Strictly speaking, $\langle 24_{\rm H}\rangle$ receives a correction after the electro-weak breaking $\langle 5_{\rm H}\rangle$ is turned on. This has to be small, and thus we omit it here; for a  discussion with the phenomenological implications for the $W$-boson mass~\cite{Senjanovic:2022zwy}, see the last part of Sec.~IV.

The renormalizable 
$d=4$ Yukawa interaction is given by
\begin{equation}
\label{eq:yukawas}
 \mathcal{L}_{\rm y}^{d=4}=\overline 5_{\rm F} Y_{\rm d} \,5_{\rm H}^* 10_{\rm F} + 10_{\rm F} \,Y_{\rm u}  \,5_{\rm H}10_{\rm F} \, ,
\end{equation}
where we ignore the generation indices, as in the rest of the paper. 
Since $\langle 5_{\rm H}\rangle$, 
{\it per se}, keeps unbroken $SU(4)$ symmetry between charged leptons and down quarks, their masses end up the same. This fails badly, ruling out the theory at the renormalizable level.

This can be easily corrected by the addition of $d=5$ Yukawa couplings~\cite{Ellis:1979fg}
\begin{equation}
\label{eq:d5yukawa}
     \mathcal{L}_{\rm y}^{d=5}= \frac{1}{\Lambda}\overline 5_{\rm F} 24_{\rm H} \,5_{\rm H}^*10_{\rm F}+...\,,
\end{equation}
where  $\Lambda$ denotes the cutoff due to some unknown new physics, and 
we assume $\Lambda\gtrsim 10 \, M_{\rm GUT}$ in order to keep the $SU(5)$ symmetry well defined. 

What happens is the following. Since $\langle 24_{\rm H}\rangle$ breaks the accidental $SU(4)$ of $\langle 5_{\rm H}\rangle$, the effective Yukawa couplings and fermion mass matrices end up being arbitrary, implying arbitrary unitary matrices that diagonalize them. This is why we stop at the leading $d=5$ term - the additional terms are unnecessary. The bottom line is that the bad predictions are gone but this has a dramatic impact on proton decay, as we will see in Sec.~IV.

It turns out that at the GUT scale, one has $m_\tau \simeq 2 \,m_b$ (see e.g.~\cite{Babu:2016bmy}) instead of them being equal as \eqref{eq:yukawas} would suggest. This implies that the cutoff cannot lie far from the unification scale - $\Lambda\lesssim 100 M_{\rm GUT}$. Remarkably, as we will see below, a similar bound emerges from the requirement of gauge coupling unification. 

The more stringent limit on the cutoff seemingly arises from neutrino mass considerations. Since neutrino mass vanishes at the renormalisable level, its leading contribution stems from the $d=5$ operator~\cite{Weinberg:1979sa}
\begin{equation}
	\label{eq:weinbergd=5}
 \mathcal{L}_{\rm \not L}=
\frac {c_{\nu}}{\Lambda} \overline 5_{\rm F}  5_{\rm H} 5_{\rm H}\overline 5_{\rm F},
\end{equation}
where $c_\nu$ is a dimensionless parameter. One obtains in turn
\begin{equation}
	\label{eq:numass}
m_\nu \simeq \frac{c_\nu}{2} \frac {v^2}{\Lambda}\,,
\end{equation}
where a factor $1/2$ arises from the renormalization from high to low energies~\cite{Babu:2001ex}.
From $m_\nu \lesssim 10^{-1}$eV, one gets 
$\Lambda \lesssim 10^{14}$GeV for $c_\nu\sim \mathcal{O}(1)$, which, at face value, would imply a too low unification scale. If one were to live on the edge of perturbativity with 
$c_\nu\sim \mathcal{O}(4 \pi)$, one would increase the limit on the cutoff scale by an order of magnitude 
$\Lambda \lesssim 10^{15}$GeV - but more about it later.

\section{Gauge Coupling Unification}
It is known that the standard model particle content alone does not suffice to unify the gauge couplings of electro-weak and strong interactions. In particular, $\alpha_1$ meets $\alpha_2$ at around $10^{12-13}\rm{GeV}$, while $\alpha_2$ and $\alpha_3$ meet much later, at 
$10^{16-17}\rm{GeV}$. In principle, though, new massive states can affect the running and potentially ensure unification. 

\begin{figure}[]
    \centering
    \includegraphics[width=0.47\textwidth]{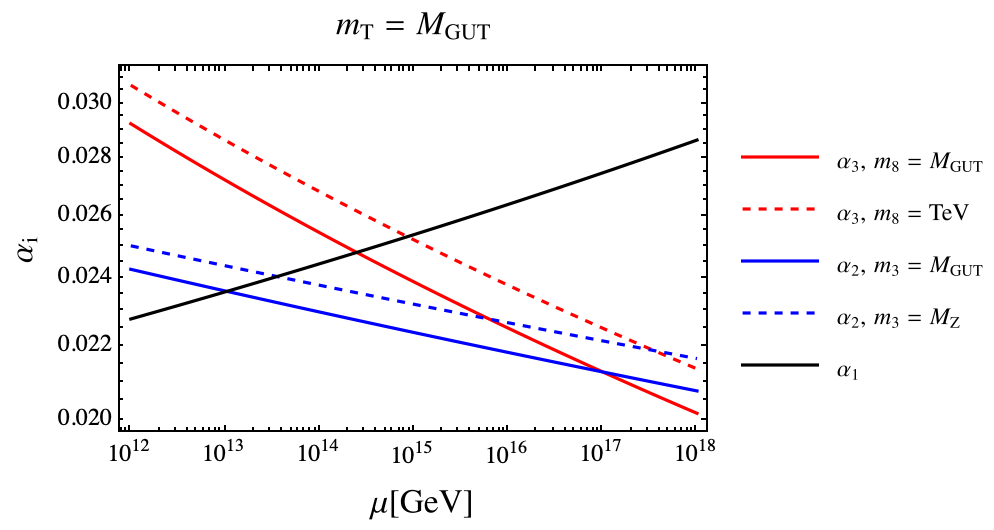}
    \caption{The failure of $m_3$ and $m_8$ at unifying the gauge couplings. }
    \label{fig:m3m8}
\end{figure}

In the minimal $SU(5)$, besides the new heavy gauge bosons which do not run below $M_{\rm GUT}$, there are only three additional states, all scalars: a weak triplet and a colour octet from $24_{\rm H}$, and a colour triplet $\rm T$ from $5_{\rm H}$. We denote their masses respectively as $m_3$, $m_8$ and $m_{\rm T}$. Since $\rm T$ induces proton decay it is customarily required to be heavier than $10^{12}\rm{GeV}$. However, it could be light, in principle, with its Yukawa couplings partially canceled by higher-dimensional operators~\cite{Dvali:1992hc}. In any case, its impact on the gauge coupling running is marginal. 

Moreover, when the dust settles (see Section~\ref{sec:4}), proton longevity requires specific mixing angles in the heavy gauge sector, thus eliminating the freedom of the Yukawa couplings of $\rm T$ - forcing it to lie close to the GUT scale. In what follows, we thus take $m_{\rm T}\simeq M_{\rm GUT}$, and vary $m_3$ and $m_8$ between $M_{\rm Z}$ and $M_{\rm GUT}$ to explore all of the phenomenologically allowed parameter space of the model.

In Fig.~\ref{fig:m3m8} it is shown explicitly how freedom of  $m_3$ and $m_8$ does not suffice to obtain unification. Note that taking $m_{\rm T}$ away from $M_{\rm GUT}$ only makes things (slightly) worse.  
It is clear that the best case scenario is to have $m_3$ ($m_8$) as small (large) as possible. 
Notice that a small $m_3$ could modify the SM $\rm W-$boson mass through an induced 
  $\langle 3_H \rangle$~\cite{Buras:1977yy}. 
The triplet decay rates are then uniquely determined by $\rm W-$mass deviation and $m_3$~\cite{Senjanovic:2022zwy,Senjanovic:2023jvv}.

Since the particle thresholds are not sufficient, one must include the additional $d=5$ gauge boson kinetic energy~\cite{Shafi:1983gz}
\begin{equation}
	\label{eq:higher}
	\Delta \mathcal{L}_{\rm kin}= \frac{c_{\rm F}}{\Lambda}\,{\rm Tr}\,F_{\mu \nu} \langle 24_{\rm H}\rangle F^{\mu \nu}\,,
\end{equation}
which modifies unification conditions to~\cite{Shafi:1983gz}
\begin{equation}
\label{eq:gaugeunif}
	\begin{split}
	\left(1-\epsilon\right)\alpha_3 (M_{\rm GUT}) &=\left( 1+\frac{3}{2} \epsilon \right)\alpha_2(M_{\rm GUT})\\
							       &= \left( 1+\frac{1}{2}\epsilon \right) \alpha_1(M_{\rm GUT})\,,
	\end{split}
\end{equation}
with the small expansion parameter  $\epsilon =c_{\rm F} M_{\rm GUT}/\Lambda$. The sign of $d=5$ term in \eqref{eq:higher} is arbitrary, and for the unification to work $\epsilon$ must be positive, so that \eqref{eq:gaugeunif} can accommodate the problem of $\alpha_1$ being bigger than $\alpha_2$ at the GUT scale, as suggested by the running. While this is achieved by choice, positive $\epsilon$ automatically ensures that the $\alpha_3$ coupling remains large enough at the unification scale. 

\begin{figure}[t]
    \centering
    \includegraphics[width=0.5\textwidth]{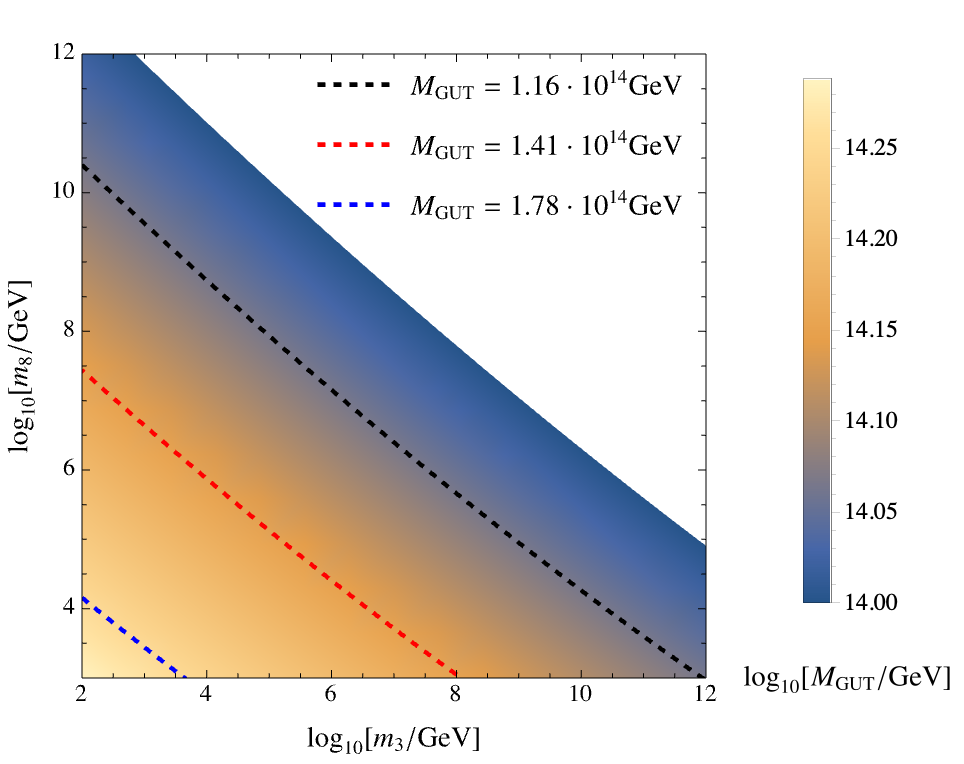}
    \caption{Dependence of $M_{\rm GUT}$ on $m_3$ and $m_8$. As discussed in the text, $m_{\rm T}\simeq M_{\rm GUT}$. }
    \label{fig:prospect}
\end{figure}

It turns out~\cite{Chakrabortty:2008zk} that the $\epsilon$ correction is sufficient {\it per se} 
(i.e., with $m_3 \simeq m_8 \simeq M_{\rm GUT}$, in accord with the survival principle which states that particle masses should lie at the largest possible scale allowed by symmetries in question~\cite{Mohapatra:1982aq}) 
to achieve unification, with $\epsilon \sim 0.04$ and $M_{\rm GUT}\simeq 5 \cdot 10^{13}\rm GeV$. It is noteworthy that this 
works precisely in the physically allowed region $10 M_{\rm GUT} \lesssim \Lambda \lesssim M_{\rm pl}/\sqrt{N}$, where the upper bond stems from the fact that $\Lambda$ can be at most as big as the perturbative gravity cutoff \cite{Dvali:2007hz,Dvali:2007wp}. Here $N$ is the number of degrees of freedom of the theory, on the order of $100$.

As shown in the next section, this would be in tension with the limits on the proton lifetime. 
However, by utilizing the freedom of $m_3$ and $m_8$, one can increase the GUT scale to actually make the theory viable. 
 This can be understood from the following simple relation
\begin{equation}
\label{eq:mgutepsilon}
\begin{split}
	\frac{M_{\rm GUT}}{M_{\rm Z}} = & \exp\left\{ \frac{\pi}{21}\left[5\left(1+\frac{\epsilon}{4} \right)\alpha_1^{-1} - 3 	\left( 1 + \frac{5\,\epsilon}{4} \right) \alpha_2^{-1}\right.\right.\\&\quad\qquad\left.\left. - 2\left( 1- \frac{5\,\epsilon}{4} \right)\alpha_3^{-1}\right]\right\}
	  \left(\frac{M_{\rm Z}^2}{m_3\, m_8} \right)^{\frac{1}{42}}\,,
\end{split}
\end{equation}
where the gauge couplings are evaluated at $M_{\rm Z}$ and $\epsilon$ turns out to be in the range $0.02 \lesssim \epsilon \lesssim 0.06$. 

It is interesting that, once unification is achieved, the color octet needs to be as light as possible in order for the unification scale to be large enough to keep the proton sufficiently long lived. In fact, for $m_3 \simeq m_8 \simeq M_{\rm Z}$, the unification scale becomes the highest, providing its absolute upper limit
\begin{equation}
\label{eq:maximalvaluemgut}
     M_{GUT} \lesssim 2\cdot 10^{14}\rm GeV\,.
\end{equation}
This can be seen explicitly in Fig.~\ref{fig:prospect}, where  
it is shown how $M_{\rm GUT}$ depends on the mass scales $m_3$ and $m_8$. For completeness, a 2-loop RG analysis was performed, taking into account proper matching conditions for thresholds~\cite{Hall:1980kf,Weinberg:1980wa}. 

The reader is probably worried that proton decay is still too fast, even for the above maximal value \eqref{eq:maximalvaluemgut} of the unification scale. However, as we show in the next section, we wish to reassure them that all is well - the freedom in fermion flavor mixing angles~\cite{Nandi:1982ew,Berezinsky:1983va,Bajc:2002bv,BPS,FileviezPerez:2004hn,Dorsner:2004xa} can keep the proton sufficiently long lived. 

\section{The flavour of proton decay}
\label{sec:4}
The proton decay in this theory stems from two different sources: the heavy ${\rm X}$ and ${\rm Y}$ gauge bosons and the new color triplet scalar ${\rm T}$, the partner of the SM Higgs doublet. As we remarked above, since ${\rm T}$ has generically small couplings to the first generation of fermions, its contribution to proton decay becomes sub-leading as soon as its mass exceeds $10^{12}$ GeV or so, and if it were on the order of the GUT scale, it would be completely negligible. 
In other words, the contribution from ${\rm T}$ can easily be suppressed, and so we concentrate on the one from the new gauge bosons. We should stress that some of our results were announced in~\cite{Senjanovic:2023jvv}.

\begin{table}[t]
\centering
\begin{tabular}{ p{4cm}p{4cm}  }
 \hline
 Channel& Lifetime ($10^{30} \rm yrs$)\\
 \hline
  $N\rightarrow e^+\, \pi$		&5300\, (n),\quad 16000\, (p) \\
  $N\rightarrow \mu^+\, \pi$		&3500\, (n),\quad 7700\, (p) \\
  $N\rightarrow \nu\, \pi$		&1100\, (n),\quad 390\, (p) \\
  $N\rightarrow e^+\, K$		&17\, (n),\quad 1000\, (p) \\
  $N\rightarrow \mu^+\, K$		&26\, (n),\quad 1600\, (p) \\
  $N\rightarrow \nu\, K$		&86\, (n),\quad 5900\, (p) \\
   \hline
\end{tabular}
\caption{Bounds on proton and neutron decay \cite{Zyla:2020zbs}.}
\label{table}
\end{table}

The effective $6-$dimensional baryon number violating operators in $SU(5)$, due to ${\rm X}$ and  ${\rm Y}$ gauge boson exchange, are schematically given by (all the fermions and anti-fermions are left-handed in our notation)
\begin{equation}
\label{eq:fermi}
	\begin{split}
		&\mathcal{L}_{\rm \not B} =\left(\overline{u}^c u\right)\left(\overline{e}^c d  + \overline{d}^c e\right) + \left(\overline{u}^c d\right) \left(\overline{e}^c u + \overline{d}^c \nu\right)\,.
	\end{split}
\end{equation}
We don't care here about the relative strength of order one between these terms. 
They induce a proton lifetime given by
\begin{equation}
\label{eq:plifetime}
	\tau_{\rm p}\simeq C\, \frac{M_{\rm GUT}^4}{\alpha_{\rm GUT}^2} m_{\rm p}^{-5},
\end{equation}
where $m_{\rm p}$ is the proton mass, $M_{\rm GUT}$ is the unification scale, $\alpha_{\rm GUT}$ is the value of gauge couplings at that scale and $C$ denotes the physical effects of running from the GUT scale to the QCD scale and the non-perturbative effects of the confinement of quarks in mesons. When the dust settles (see Appendix), one typically claims $C \simeq 1$.  
The most stringent bound on proton lifetime from Super-Kamiokande is about $\tau_{\rm p}\gtrsim10^{34}\rm{yrs}$ for the channel ${\rm p }\rightarrow \pi^0\, e^+$~\cite{Super-Kamiokande:2020wjk}. For $\alpha_{\rm GUT}\simeq40^{-1}$ (obtained from the running), and if $C$ were really of order one, one would get
$M_{\rm GUT}\gtrsim 4\cdot 10^{15}\rm{GeV}$
which would be too high to be compatible with the values obtained in the previous section. 

This, however, is not enough to rule out minimal $SU(5)$, not before studying carefully the value of $C$, 
which turns out to be intimately connected with the flavour structure of the above baryon decay operators. What happens is that $C$ could in principle be quite suppressed by small flavor mixings.
This was studied carefully in Ref.~\cite{Dorsner:2004xa} and we briefly recapitulate here their arguments and update their results.

The crucial point, as we discussed in the Sec.~II, is that due to the presence of higher-dimensional operators, Yukawa couplings become arbitrary,
which implies arbitrary fermion mass matrices and, in turn, arbitrary unitary transformations from the weak to the physical mass eigenstates base. 

Rotating the fermions into the mass basis, \eqref{eq:fermi} becomes 
\begin{equation}
\label{eq:fermi_mass}
	\begin{split}
		&\mathcal{L}_{\rm \not B} =
  \left(\overline{u}^c\, U_c^\dagger U\, u\right)\left(\overline{e}^c\, E_c^\dagger D\,d + 	\overline{d}^c\, D_c^{\dagger}E\,e\right)\\
		& + 	\left(\overline{u}^c\, U_c^\dagger D \,d\right)\left(\overline{e}^c\, E_c^{\dagger}U\,u + 
		\overline{d}^c\, D_c^{\dagger}N\,\nu \right)\,.
	\end{split}
\end{equation}
The capital symbols denote the unitary transformations 
\begin{equation}
\label{eq:rotations}
f \to F\,f \,; \,\,\,\,\, f^c \to F_c\, f^c \,; \,\,\,\,\, F F^\dagger = F_c F_c\,^\dagger = 1\, ,
\end{equation}
needed to diagonalize fermion mass matrices in the obvious notation ($N$ being rotation on neutrinos). Recall that the CKM matrix is given by $V_{\rm CKM}= U^{\dagger}D$. 

\subsection{Analytical study: preliminaries}
One could even imagine rotating proton decay completely away, however, unitarity forbids it. In fact, for this to happen one would need~\cite{Nandi:1982ew}
\begin{equation}
\label{eq:pdecayconditions}
	\begin{split}
		&\left(U_c^\dagger U\right)_{11}=0\,,\,\,\,\,\,\,
		\left(E_c^\dagger U\right)_{a1}=0\,,\,\,\,\,\,\,
		\left(U_c^\dagger D\right)_{1a}=0\,,
	\end{split}
\end{equation}
for $a=1,2$. 
It is readily seen that the third equations in \eqref{eq:pdecayconditions} cannot be simultaneously satisfied for both $a=1,2$ as long as $( V_{\rm CKM})_{13}\neq 0$.  
Still, one can exploit the available freedom in these mixings in order to maximize the proton decay lifetime. 

In order to set the stage, we report in Table~\ref{table} the experimental bounds on proton and neutron decays~\cite{Zyla:2020zbs} relevant for the analysis. We are best off with killing the pion and positron or anti-muon modes, in order to try to keep the GUT scale as low as possible and salvage the theory. 
With that in mind, a natural thing to try would be $\left(U_c^\dagger U\right)_{11}=0,\,\left(E_c^\dagger U\right)_{a1}=0,\,\left(U_c^\dagger D\right)_{11}=0$ for $a=1,2$. In this case, proton (neutron) could decay into pions and kaons, and antineutrino and the decay rates would be suppressed by small $( V_{\rm CKM})_{13}$. In this case, $M_{\rm GUT}\geq \sqrt{\alpha_{GUT}/40^{-1}}\,3 \cdot 10^{14}\rm GeV$, too large to be  compatible with gauge coupling unification, c.f.,~\eqref{eq:maximalvaluemgut}. 
		 
A better choice is
		\begin{equation}
  \label{eq:case2}
    \begin{split}
	\left(U_c^\dagger D\right)_{1a}=&
	\left(E_c^\dagger D\right)_{1a}=\left(E_c^\dagger              D\right)_{a1} = 0\,,\\
	&\left(D_c^\dagger E\right)_{1a}= \left(D_c^\dagger        E\right)_{a1} =0\,,
    \end{split}
\end{equation}
for $a=1,2$. 
This leads to the following lower bound on the GUT scale\footnote{Chiral Lagrangian parameters were taken from~\cite{Yoo:2021gql}. 
}
\begin{equation}
\label{eq:bound}
M_{\rm GUT} \geq \sqrt{\frac{\alpha_{\rm GUT}}{40^{-1}}} 
      1.3\cdot 10^{14}\rm{GeV}  \,.
\end{equation}
Here the relevant channel is $p \rightarrow K^0 + \mu^+$, again with $(V_{\rm CKM})_{13}$ suppression. Table~\ref{table} shows that this mode is not so well measured, which helps the theory pass the unification test~\eqref{eq:maximalvaluemgut}.
 
Fig.~\ref{fig:prospect} shows that \eqref{eq:bound} implies $m_3,m_8\lesssim 10^{8}\,\rm GeV$.
Moreover, an improvement in $K^0$ and $\mu^+$ lifetime by a factor of about 4-5 would require $m_3$ and $m_8$ below $10\,\rm TeV$ - in the right ballpark for hadron colliders. 

It is noteworthy that \eqref{eq:bound} differs from analogous result of~\cite{Dorsner:2004xa} by roughly a factor of $2$. Over the years, both the nucleon lifetime limits and the chiral Lagrangian parameters changed, which could explain the discrepancy. Furthermore,~\cite{Dorsner:2004xa} seems to have omitted the running of the proton decay amplitude 
from the weak to the unification scale.   

\subsection{Numerical study: maximizing proton decay rotation}
Naively, it would seem that the theory favours particular channels mentioned above, and moreover, that it could soon be ruled out. The trouble is that the zeroes in mixing matrices need not be exact, a small leakage would allow for other decay rates, such as the pion and charged anti-leptons one.
In order to ensure a complete exhaustion of the parameter space, we performed a Monte Carlo exploration - whose details can be found in Appendix - in which the rotation conditions were allowed to be more general. 

\begin{figure}[]
    \centering
    \includegraphics[width=0.47\textwidth]{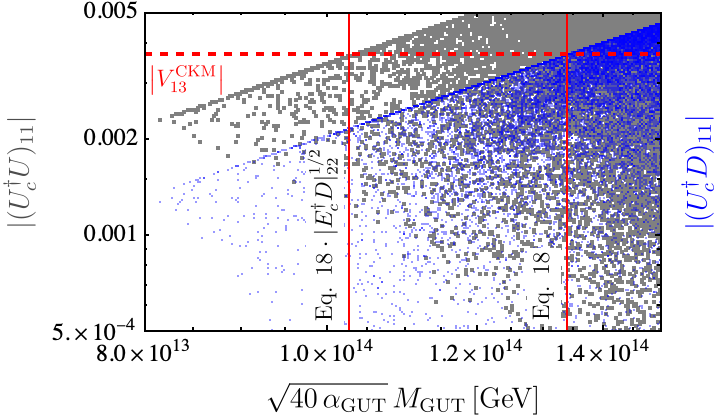}
    \caption{Randomized values of $|(U_c^\dagger U)_{11}|$ (gray) and corresponding $|(U_c^\dagger D)_{11}|$ (blue) vs $M_{\rm GUT}$ compatible with experimental decay rate bounds. Dashed red line corresponds to $|(V_{\rm CKM})_{13}|$. The red left line corresponds to $|E_c^\dagger D|_{22}=1$ leading to Eq.~\eqref{eq:bound}, while the red right line corresponds to its relaxation discussed in the appendix, c.f.~\eqref{eq:ecdrel}.}
    \label{fig:matrixelements}
\end{figure}

The resulting analysis is summarized in Fig.~\ref{fig:matrixelements}, in which  the absolute values of $(U_c^\dagger U)_{11}$ (gray) and the corresponding $(U_c^\dagger D)_{11}$ (blue) are shown as a function of $M_{\rm GUT}$. A similar plot holds for the $(U_c^\dagger D)_{12}$ matrix element, which we omit here. 
The result of Eq.~\eqref{eq:bound}, is automatically found by the numerical analysis, 
which is shown with the vertical red line in Fig.~\ref{fig:matrixelements}.
All points to its left - with matrix elements whose values are smaller than $(V_{\rm CKM})_{13}$ - correspond to effective realizations compatible with experimental limits on nucleon lifetimes with a lower $M_{\rm GUT}$.

It is not surprising that such points correspond to matrix elements whose absolute values are below $|(V_{\rm CKM})_{13}|$, as in the case of $|(U_c^\dagger D)_{11}|$. 
The lower bound on $M_{\rm GUT}$ compatible with current proton decay lifetime is then found to be
\begin{equation}
\label{lowestbound}
    M_{\rm GUT}\gtrsim \sqrt{\frac{\alpha_{\rm GUT}}{40^{-1}}}\,8\cdot 10^{13}\,\rm GeV.
\end{equation}
Due to QCD low-energy inputs, affected by uncertainties (see, e.g.,~\cite{Yoo:2021gql}), our results are numerically reliable up to corrections of about $10\%$.
Eq.~\eqref{lowestbound} shows that the rotation of all nucleon decay amplitudes into $(V_{\rm CKM})_{13}$, discussed above, does not  exhaust all of the parameter space. 

If the reader feels that there is something unnatural in having a theory live on what looks like a convoluted point, we remind them of the way the GIM suppression of $K_L-K_S$ works as a conspiracy of a small charm quark mass, and then a huge top mass contribution suppressed by a tiny $(V_{\rm CKM})_{13}$. Good theories live dangerously.
 
In our analysis we exploited the freedom in different proton decay channels, in order to obtain the lowest possible unification scale compatible with proton lifetime. Therefore, an optimistic scenario of an improvement in proton lifetime by a factor of $10-25$ in future experiments for all channels, would still not rule out the theory. It would, however, indicate the presence of either $m_3$ or $m_8$ below $100\,\rm TeV$. An ulterior improvement roughly up to a factor $25-50$ in proton lifetime would require these particles to be lighter than $10\,\rm TeV$. The principal uncertainty in these numbers stems from the QCD computation of the nucleon decay matrix elements~\cite{Yoo:2021gql}.
In fact, a $10\%$ uncertainty on the $M_{\rm GUT}$ bound, implies, roughly, a $50\%$ uncertainty in the necessary improvement on proton lifetime.

The necessary flavor rotation in heavy gauge boson interactions implies, however, the impossibility of doing the same for the Yukawa interactions of the $\rm T$ scalar, forcing it to lie roughly close to the GUT scale.

\,

\noindent\textbf{Consequences for neutrino mass. }As we discusses in the Sec.~II, when defining the minimal realistic $SU(5)$, one resorts to higher dimensional operators, implying that neutrino mass ought to be given by \eqref{eq:numass}. The crucial question here is how perturbative one has to be. If one stays at its boundary, one gets $\Lambda\sim 10 M_{\rm GUT}\sim 10^{15}\,\rm GeV$ and all is well. If one wanted, though, to keep the couplings at worst of order one, this would require $\Lambda\lesssim 10^{14}\,\rm GeV$, which would seemingly invalidate the theory. In such a case the reader may wonder why we we would go through all of this tedium, if the theory was already ruled out by neutrino mass consideration. We wanted to show, however, that the often repeated claim of the minimal $SU(5)$ not being able to provide the unification of gauge couplings with the sufficient proton longevity is simply wrong. There is more to it, though.

The point is that most completions of the $d=5$ operator in \eqref{eq:weinbergd=5}  tend to be based on new fermions or scalars, with potentially small Yukawa couplings. The most popular example of this realization is the seesaw mechanism, which in its original form consists of the addition of RH neutrino singlets. Even if they were much lighter than the GUT scale, and even if they were at the weak scale itself, there would be almost no change of the original theory due to the associated smallness of their Yukawa couplings - except for the emerging neutrino mass. In other words, the physical cutoff scale in \eqref{eq:numass} could be arbitrarily small without affecting the unification scale and proton decay.


\section{SU(5) extensions and neutrino mass}

 In principle, the minimal theory could be defined even with the addition of RH neutrinos, due do their inoffensive nature. There is something rather unappealing, though, in adding gauge singlets in a theory whose predictions depend profoundly on its gauge nature. There are two particularly attractive alternatives, based on the so-called type II and type III seesaw mechanisms (for a review, see e.g.~\cite{Senjanovic:2011zz}), with minimal changes of the theory and possibly predictive outcomes, both obtained by an addition of a single field augmented with higher-dimensional operators as in the original minimal theory. We recapitulate here their salient features in view of our findings.

\paragraph{Type II seesaw}
 Instead of fermion singlets, the RH neutrinos, one may opt for the so-called type II seesaw~\cite{Magg:1980ut,Mohapatra:1980yp,Lazarides:1980nt} through the addition of a symmetric scalar $15_{\rm H}$ field~\cite{Glashow:1979nm,Senjanovic:1981ff}. This induces the following new relevant couplings
\begin{equation}
\label{eq:yukawaII}
 \mathcal{L}_{type\, II}=\overline 5_{\rm F} Y_\nu 15_{\rm H} 10_{\rm F} + \mu \,5_{\rm H} \,15_{\rm H}^*\, 5_{\rm H}+...\,,
\end{equation}
which gives neutrino a Majorana mass through the small vev of $15_{\rm H}$.
 While it may be considered somewhat ad-hoc, an outcome of model building, it has an appeal of being potentially verifiable as the origin of neutrino mass. Namely, the weak triplet in $15_{\rm H}$ contains doubly charged scalars, whose branching ratio into charged leptons are dictated by the neutrino mass matrix~\cite{Garayoa:2007fw,Kadastik:2007yd} - of course, assuming these scalars can be produced at hadron colliders.   
The minimal $SU(5)$ model is a sub-case of this one, and therefore there exists a region of the parameter space where all of the sub-multiplet of $15_{\rm H}$ has a mass around the unification scale.

The problem is that the states in $15_{\rm H}$ do not suffice to sufficiently increase the GUT scale~\cite{Dorsner:2005fq}, therefore still requiring cancellations in proton decay amplitudes. In other words, from the point of view of constructing a minimal realistic theory, one might as well stick to the original theory, as we amply discussed above.

\paragraph{Type III seesaw}  Another simple alternative is based on a fermionic $24_{\rm F}$ field~\cite{Bajc:2006ia,Bajc:2007zf} instead of $15_{\rm H}$. A naive analysis which ignores \eqref{eq:higher} and the flavour of proton decay would imply the presence of the weak triplet fermions and scalars, within the reach of the LHC~\cite{Bajc:2006ia} - an exciting prediction, if true (the detailed study of the resulting phenomenology can be found in~\cite{Arhrib:2009mz}). The fermion triplet variation of the seesaw mechanism is coined type III~\cite{Foot:1988aq}, with today's experimental limit on the triplet mass being roughly TeV~\cite{Novak:2020jju}.

However, the additional freedom of Eq.~\eqref{eq:higher} leads to a straighforward generalization of
 \eqref{eq:mgutepsilon}
\begin{equation}
\label{eq:mgut24f}
	\begin{split}
	\frac{M_{\rm GUT}}{M_{\rm Z}} =& e^{ \frac{\pi}{21}\left[5\left(1+\frac{1}{4}\epsilon \right)\alpha_1^{-1} - 3 	\left( 1 + \frac{5}{4}\epsilon \right) \alpha_2^{-1} - 2\left( 1- \frac{5}{4}\epsilon \right)\alpha_3^{-1}\right]}\\
	&\quad\quad \quad \left(\frac{m_{\rm LQ}^4\,M_{\rm Z}}{m_3^{1/2}\, m_8^{1/2} m_{3\rm F}^2 m_{8\rm F}^2} \right)^{\frac{1}{21}\left(1 - \frac{1}{4}\epsilon\right)},
\end{split}
\end{equation} 
where the labels $3 \rm F$, $8\rm F$ and $\rm LQ$ denote respectively the weak triplet, the colour octet and the leptoquark component of $24_{\rm F}$. 
In the previous section, a low $m_3$ and $m_8$ could increase unification up to $10^{14}\,\rm{GeV}$. It is not surprising then that lowering also the masses of their fermionic counterpart can give $M_{\rm GUT}\gtrsim 10^{15}\,\rm GeV$. 

While for $\epsilon =0$, both $m_{3\rm F}$ and $m_3$ lie close to TeV energies~\cite{Bajc:2006ia}, it is a straightforward numerical exercise to show that for $\epsilon \simeq 0.05$, all the particle masses are above $10^{5}\rm GeV$, with a unification scale of $4\cdot 10^{15}\,\rm GeV$, compatible with the naive proton lifetime estimate. 

A comment is noteworthy here. Ref.~\cite{Bajc:2006ia} made two assumptions: (i) no flavour cancellations in proton decay amplitudes and (ii) that the cutoff is as large as possible, in order to ensure precise perturbative predictions. The latter implies $\Lambda \simeq 100 \,M_{\rm GUT}$ or $\epsilon \simeq 0.01$ (a larger cutoff would not suffice to correct the bottom-tau mass relation), which then justifies the predictions quoted. 

A similar situation emerges also in minimal $SO(10)$ theory with higher dimensional operators~\cite{Preda:2022izo}. In fact, also there the presence of new light particle states is intimately tied to the assumption of no cancellations in proton decay amplitudes. It is precisely in this sense that the minimal $SU(5)$ theory stands - it is predictive without any further assumption.

\section{Summary and outlook}

In the last twenty years or so, people - including one of the authors - have been regularly writing necrologies for the the minimal $SU(5)$ theory, even in its non-renormalisable version. And so, people - now including both authors - have been arguing in favor of going beyond the minimal theory.  
As we showed here, in spite of these morbid desires, the minimal $SU(5)$ theory is alive and kicking - as long as one is willing to accept higher-dimensional operators as the simultaneous cure of the wrong fermion mass relations and the failure of the gauge coupling unification. However, it does live dangerously and an improvement in proton lifetime bounds could require new light scalar states, the weak triplet and the colour octet. This strengthens a case for proton decay searches and future hadron colliders. 

Moreover, a light scalar weak triplet  would be expected to change the Standard Model W-mass relation, paving a window for an indirect observation of new physics. Remarkably, its low energy effective theory is unambiguously determined by its mass and the W-mass deviation~\cite{Senjanovic:2022zwy}.  

Admittedly, higher-dimensional operators may not be that appealing, but the vicinity of the non-perturbative gravitational effects to the GUT scale makes them  potentially sizeable  and thus hard to ignore. 
In this minimal scenario they are actually a must, being the only possible source of neutrino masses through the $d=5$ Weinberg operator. 
Moreover, the actual scale of this operator could be independent of the cutoff and naturally much lower, due to small Yukawa couplings of particles that provide UV completions of the theory. 

Otherwise, in order to allow for a higher cutoff scale, one would have to add fermionic singlets (the RH neutrinos), an admittedly ugly solution. For this reason we discussed
more natural extensions that utilize  
the inclusion of $15_{\rm H}$ or $24_{\rm F}$ states~\cite{Dorsner:2005fq,Bajc:2006ia} This allows for possible direct collider probes of the neutrino mass origin - an exciting possibility - but there is no guarantee to have the new particles accessible at the LHC or a new hadron collider. The case of the adjoint fermion $24_{\rm F}$ is free from a necessity of suppressing the proton decay by tuning the flavour structure and at the first glance seemed to predict a weak fermion triplet at TeV energies. The higher-dimensional operators, though, in the gauge field sector could push it above  $100\, \rm TeV$, and render it inaccessible even at the next hadron collider. While this model remains a natural extension of the minimal theory, our results show that it is less predictive than imagined originally.

We wish to end on a more sober note regarding the whole idea of grand unification. The main message of our work, besides the fact that the minimal $SU(5)$ theory with higher dimensional operators is still very much viable, is that grand unified theories have a generic problem of a lack of clear predictions when it comes to nucleon decay. But nucleon decay is the essence of grand unification, just as the neutral currents were the essence of the SM. 
The main problem is the fact that the GUT scale is astronomically large compared to the weak one, and one is often forced to make assumptions - non testable in any foreseeable future - at $M_{\rm GUT}$ in order to claim verifiable consequences at accessible energies. However, it is precisely the enormity of the GUT scale that allows for a successful effective $d=6$ 
theory of baryon violation~\cite{Weinberg:1979sa,Wilczek:1979hc}, with a number of clear predictions (for an overview, see~\cite{Senjanovic:2009kr}) that hold true if $M_{\rm GUT} \gg M_{\rm Z}$. 

Establishing the nature of the effective theory of baryon number violation through experiment would be equivalent to the historic determination of the V-A character of the effective Fermi theory of weak interaction - which then allowed to create the fundamental renormalisable theory, the Standard Model. As Weinberg put it nicely~\cite{Weinberg:2009zz}: ``V-A was the key". One perhaps ought to wait for an analogous thing to happen in the nucleon decay interaction in order to find a fundamental grand unified theory?

\,

\paragraph*{\textbf{Acknowledgements.}} We are grateful to Anca Preda for participating at the initial stages of this projects. We are indebted to Borut Bajc for important comments and a careful reading of the manuscript, and to Kaladi Babu for a correspondence regarding proton decay rates. We also wish to thank Alejandra Melfo for her help in improving the presentation of our work. 

\,

\appendix
\section*{Appendix: Decay Rate analysis}
This appendix relaxes conditions \eqref{eq:bound}. Namely, we look for a lower bound on $M_{\rm GUT}$ by allowing for arbitrary matrix elements in the $d=6$ effective operators \eqref{eq:fermi_mass}. To do so, a numerical sampling of the parameter space has been performed via a Monte Carlo method that will be explained below. For simplicity and completeness, the notation of Ref.~\cite{FileviezPerez:2004hn} will be adopted and adapted to the following.

\subsection{Matrix Elements}
The $d=6$ operators leading to proton decay in $SU(5)$ grand unification can be derived from integrating out the heavy gauge bosons $X,Y$ with electric charge $4/3$ and $1/3$ respectively. The resulting Standard Model operators are $B-L$ conserving and are given by~\cite{Weinberg:1979sa,Weinberg:1980bf,Wilczek:1979hc,Senjanovic:2009kr}
\begin{eqnarray}
\label{O1}
\textit{O}^{B-L}_I&=& \frac{g^2}{2M_X^2}
\ \epsilon_{ijk} \ \epsilon_{\alpha \beta} 
\ \overline{u_{i a}^C} \ \gamma^{\mu} \ Q_{j \alpha a}   \
\overline{e_b^C} \ \gamma_{\mu} \ Q_{k \beta b}\\ 
\label{O2}
\textit{O}^{B-L}_{II}&=& \frac{g^2}{2M_X^2}
\ \epsilon_{ijk} \ \epsilon_{\alpha \beta}
\ \overline{u_{i a}^C} \ \gamma^{\mu} \ Q_{j \alpha a}   \
\overline{d^C_{k b}} \ \gamma_{\mu} \ L_{\beta b}\qquad
\end{eqnarray}
where $g$ denotes the gauge coupling at $M_X\simeq M_{\rm GUT}$ and $Q= ( u, d)$, $L= ( \nu, e)$; 
$i$, $j$ and $k$ are the color indices, $a$,$b$ are the family
indices, and $\alpha, \beta =1,2$. 
\\
When going to the physical basis, the effective operators~(\ref{O1},\ref{O2}) become
\begin{eqnarray}
\label{Oec}
&\textit{O}&(e_{\alpha}^C, d_{\beta})= \frac{g^2}{2M_X^2} c(e^C_{\alpha},
d_{\beta})  \epsilon_{ijk} 
 \overline{u^C_i}  \gamma^{\mu}  u_j  \overline{e^C_{\alpha}} 
\gamma_{\mu}  d_{k \beta}\,\,\, \\
\label{Oe}
&\textit{O}&(e_{\alpha}, d^C_{\beta})= \frac{g^2}{2M_X^2} c(e_{\alpha}, d^C_{\beta})  \epsilon_{ijk}  
\overline{u^C_i}  \gamma^{\mu}  u_j  \overline{d^C_{k \beta}} 
\gamma_{\mu}  e_{\alpha}\\
\label{On}
&\textit{O}&(\nu_l, d_{\alpha}, d^C_{\beta} )= \frac{g^2}{2M_X^2} c(\nu_l, d_{\alpha}, d^C_{\beta}) 
\epsilon_{ijk}  \overline{u^C_i}  \gamma^{\mu}  d_{j \alpha}
 \overline{d^C_{k \beta}}  \gamma_{\mu}  \nu_l \qquad\,\,
\end{eqnarray}
where
\begin{eqnarray}
\label{cec}
&c&(e^C_{\alpha}, d_{\beta})= \left(U_c^\dagger U\right)_{11} \left(E_c^\dagger D\right)_{\alpha \beta} + ( U_c^\dagger D)_{1
\beta}( E_c^\dagger U)_{\alpha 1}\qquad \\
\label{ce}
&c&(e_{\alpha}, d_{\beta}^C) =   \left(U_c^\dagger U\right)_{11}\left(D_c^\dagger E\right)_{\beta \alpha} \\
\label{cnu} 
&c&(\nu_l, d_{\alpha}, d^C_{\beta})=  \left( U_c^\dagger D \right)_{1 \alpha}
(D_c^\dagger N)_{\beta l} \qquad
\end{eqnarray}
This is the precise form of $\eqref{eq:fermi_mass}$.

\subsection{Two body decay channels of the nucleon}
In proton decay experiments, no distinction is made between the flavour of neutrino, nor the chirality of the charged lepton in the decay products. In this case, the relevant decay channel can be derived with the appropriate chiral techniques~\cite{Claudson:1981gh,Chadha:1983sj,FileviezPerez:2004hn} and are given by
\begin{widetext}
\begin{eqnarray}
\label{A1}
\Gamma(p \to K^+\bar{\nu})
		&=& \frac{g^4}{4M_{\rm X}^4}\frac{(m_p^2-m_K^2)^2}{8\pi m_p^3 f_{\pi}^2} A_L^2 
\left|\alpha\right|^2
\sum_{i=1}^3 \left|\frac{2m_p}{3m_B}D \ c(\nu_i, d, s^C) 
+ [1+\frac{m_p}{3m_B}(D+3F)] c(\nu_i,s, d^C)\right|^2\\
\label{A2}
\Gamma(p \to \pi^+\bar{\nu})
		&=& \frac{g^4}{4M_{\rm X}^4}\frac{m_p}{8\pi f_{\pi}^2}  A_L^2 \left|\alpha
		\right|^2 (1+D+F)^2 
\sum_{i=1}^3 \left| c(\nu_i, d, d^C) \right|^2\\
\label{A3}
\Gamma(p \to \eta e_{\beta}^+) 
		&=& \frac{g^4}{4M_{\rm X}^4} {(m_p^2-m_\eta^2)^2\over 48 \pi f_\pi^2 m_p^3}
A_L^2 \left|\alpha \right|^2 (1+D-3 F)^2 \{ \left| c(e_{\beta},d^C)\right|^2 +\left|c(e^C_{\beta}, d)\right|^2 \}\\
\label{A4}
\Gamma (p \to K^0 e_{\beta}^+) 
		&=& \frac{g^4}{4M_{\rm X}^4} {(m_p^2-m_K^2)^2\over 8 \pi f_\pi^2 m_p^3}  A_L^2
		\left|\alpha\right|^2 [1+{m_p\over m_B} (D-F)]^2 
\{ \left|c(e_{\beta},s^C)\right|^2 +  \left|c(e^C_{\beta},s)\right|^2\}\\
\label{A5}
\Gamma(p \rightarrow \pi^0 e_{\beta}^+)
           &=& \frac{g^4}{4M_{\rm X}^4} \frac{m_p}{16\pi f_{\pi}^2} A_L^2 \left|\alpha\right|^2
		(1+D+F)^2 \{ \left|c(e_{\beta},d^C)\right|^2 + 
		\left|c(e^C_{\beta},d)\right|^2 \}\\
\label{A6}
\Gamma(n \to K^0 \overline\nu)&=&  \frac{g^4}{4M_{\rm X}^4}
\frac{(m_n^2-m_K^2)^2}{8 \pi m_n^3 f_\pi^2} A_L^2 \left|\alpha\right|^2 \cdot
\nonumber\\ &&\cdot  \sum_{i=1}^3 \left|c(\nu_i,d,s^C) [1+\frac{m_n}{3 m_B} (D-3 F)]-c(\nu_i,s,d^C)[1+\frac{
		m_n}{3 m_B}(D+3 F)]\right|^2\\
\label{A7}
\Gamma(n \to \pi^0 \overline\nu)&=& \frac{g^4}{4M_{\rm X}^4}\frac{m_n}{16 \pi f_\pi^2} A_L^2
		\left|\alpha\right|^2 (1+D+F)^2 \sum_{i=1}^3 \left|c(\nu_i, d,
		d^C)\right|^2\\
\label{A8}
\Gamma(n \to \eta \overline\nu)&=& \frac{g^4}{4M_{\rm X}^4} \frac{(m_n^2-m_\eta^2)^2}{48 \pi m_n^3
f_\pi^2} A_L^2 \left|\alpha\right|^2 (1+D-3 F)^2 \sum_{i=1}^3 \left|c(\nu_i, d, d^C)\right|^2\\
\label{A9}
\Gamma(n \to \pi^- e^+_{\beta})&=& \frac{g^4}{4M_{\rm X}^4}\frac{m_n}{8 \pi f_\pi^2} A_L^2
		\left|\alpha\right|^2 (1+D+F)^2
		\{\left|c(e_{\beta}, d^C)\right|^2 +  \left|c(e^C_{\beta},d)\right|^2\}
\end{eqnarray} 
\end{widetext}
where the average baryon mass $ m_B \approx
m_\Sigma \approx m_\Lambda\simeq 1.15\, \rm GeV$, $D\simeq0.8$, $F\simeq0.47$ and $\alpha\simeq0.012\,\rm GeV^3$ are the parameters
of the chiral lagrangian taken from~\cite{Yoo:2021gql}. 
Here all coefficients of four-fermion operators are evaluated at
$M_Z$ scale. $A_L$ takes into account renormalization effects
from $M_Z$ to
GeV - corresponding to a factor of about $1.4$~\cite{FileviezPerez:2004hn}. A further correction arises due to the gauge coupling running from $M_{\rm GUT}$ to $M_Z$~\cite{Abbott:1980zj} which gives an additional factor 3.4. 
Finally, $\nu_i= \nu_e, \nu_{\mu}, \nu_{\tau}$ and $e_{\beta}= e, \mu$. 

Given the above values, direct inspection of \eqref{A5} fixes the prefactor coefficient $C\simeq 0.73$ in \eqref{eq:plifetime}, as mentioned in the main text. 

\subsection{On the numerical analysis}
\label{moreonnumerics}
The most stringent bound found by Ref.~\cite{Dorsner:2004xa} is obtained with
\begin{equation}
\begin{split}
\label{eq:case22}
			&\left(U_c^\dagger D\right)_{1a}=0\\
			&\left(E_c^\dagger D\right)_{1a}=\left(E_c^\dagger D\right)_{a1} = 0,\\
			&\left(D_c^\dagger E\right)_{1a}= \left(D_c^\dagger E\right)_{a1} =0,\\
                & \left(E_c^\dagger U\right)_{a1}=0
			\end{split}
\end{equation}
for $a=1,2$. 
The first condition of Eq.~\eqref{eq:case22} immediately leads to 
\begin{equation}
\label{eq:minres}
|\left(U_c^\dagger U\right)_{11}|=|\left(V_{\rm CKM}\right)_{13}|.
\end{equation}

Second and third condition in Eq.~\eqref{eq:case22} cannot be accommodated for all four matrix elements. In particular, bound~\eqref{eq:bound} is obtained by setting all matrix elements to vanish, except for the $|\left(E_c^\dagger D\right)_{22}|=1$. This effectively suppresses the decay into $K\mu$ by $(V_{\rm CKM})_{13}$. 

Indeed, a less stringent condition can be imposed. Namely, all the elements $(E_{c}^\dagger D)_{ab}$, $a,b=1,2$ can be chosen to minimize equally proton decay into the different charged lepton channels. 
An inspection of (\ref{cec},\ref{ce}), suggests it would be natural to take $|(E_c^\dagger D)| = |(D_c^\dagger E)^T|$. Given current values of proton lifetime~\cite{Zyla:2020zbs}, the $2\times 2$ block matrix in the first two families minimizing decay rates in Table~\ref{table} - while preserving unitarity of the total $3\times 3$ matrix - is given by
\begin{equation}
\label{eq:ecdrel}
     |(E_c^\dagger D)|\simeq |(D_c^\dagger E)^T|\simeq 
    \begin{pmatrix}
   0.11 &0.77\\
0.16 &0.61
    \end{pmatrix}.
\end{equation}
In order to relax the first condition in~\eqref{eq:case22}, leading to~\eqref{eq:minres}, we numerically randomized the choice of $(U_c^\dagger U)_{11}$ and used $U^\dagger D=K_1 V_{\rm CKM}K_2$ (where $K_{1,2}$ are two diagonal matrices with respectively three and two phases) to compute the matrix $U_c^\dagger D$. Each sampled solution was then used to derive the corresponding $M_{\rm GUT}$.

\bibliography{biblio}

\end{document}